\documentclass[12pt]{iopart}
\usepackage{graphicx}
\usepackage{amssymb}

\begin{document}

\title{A study of the electromagnetic fluctuation induced forces on thin metallic films}
\date{\today}

\author{A.~Benassi$^{1,2}$, C.~Calandra$^2$}

\address{$^1$ CNR/INFM-National Research Center on nanoStructures and bioSystems at Surfaces
(S3), Via Campi 213/A, I-41100 Modena, Italy}
\address{$^2$ Dipartimento di Fisica, Universit\`a di Modena e Reggio Emilia, Via Campi 213/A, I-41100 Modena,
Italy} \ead{benassi.andrea@unimore.it}

\begin{abstract}
Using the plasma model for the metal dielectric function we have calculated the electromagnetic fluctuation induced forces on a free standing metallic film in vacuum  as a function of the film size and the plasma frequency. The force for unit area  is attractive and for a given film thickness it shows an intensity maximum at a specific plasma value, which cannot be predicted on the basis of a non retarded description of the electromagnetic interaction. If the film is deposited on a substrate or interacts with a plate, both the sign and the value of the force are modified. It is shown that the force can change sign from attraction to repulsion upon changing the substrate plasma frequency. A detailed comparison between the force on the film boundaries and the force between film and substrate indicates that, for $50-100nm$ thick films, they are comparable when film-substrate distance is of the order of the film thickness.
\end{abstract}


\section{Introduction}
\label{introduction} Electromagnetic fluctuation induced forces
have been the subject of several investigations both in the non
retarded small distance limit (dispersion forces) and in the
retarded large distance case (Casimir
forces)\cite{lifshitz,dzyaloshinskii,mahanty,milloni,bordag,lamoreaux}.
Since the basic work from Lifshitz\cite{lifshitz}, studies have been focused mainly on the determination of the 
forces between two semi-infinite planar media\cite{dzyaloshinskii} or between a sphere and a planar medium\cite{derjaguin}.
Even if model calculations have been performed for special
geometries\cite{bordag,lamoreaux,barton}, forces on realistic
systems have been
obtained starting from the above mentioned configurations.\\
For several technological applications, multilayer systems, obtained by depositing thin films of different
species onto a given substrate, are important and theoretical approaches have been devised to determine van der Waals
forces between laminated media\cite{podgornik,white}.
Casimir forces between moving parts have been considered as possible source of instabilities in micro and
nano-devices, where the components are in close proximity\cite{serry,buks,zhao}. In many of these systems the
situations of interest correspond to interaction between parallel interfaces of films and plates with micro
or submicro-size and submicro-distances. While some of these studies have been performed using simplified
models for the interaction, like the assumption that the interaction is correctly represented by the force
between ideal metallic plates, it has been pointed out that a realistic description of adhesion or stiction
phenomena has to account for the dependence of the force upon the shape and optical properties of the
components\cite{white,barcenas}.\\
In spite of this large amount of work, a detailed study of the
behaviour of the electromagnetic fluctuation induced forces in
unsupported or deposited conductive films, as a function of their
size and optical parameters has not been published. The interest
has been focused mainly on the interaction between two-dimensional
films, for which forces are supposed to show a peculiar dependence
upon the film distance\cite{sernelius,boestrom1,boestrom2}. Less
interest has been given to the study of the forces on the film
boundaries due to vacuum fluctuations, which are present even in
an isolated film and depend upon its size and properties.\\
We can formulate the problem as follows: suppose we have a simple
metal, whose dielectric function can be expressed as
\begin{equation}
\epsilon(\omega)=1-\frac{\Omega_{p}^{2}}{\omega(\omega+i/\tau)}
\label{prima}
\end{equation}
where $\Omega_{p}$ is the  plasma frequency and $\tau$ is the relaxation time. If we consider an
isolated metal film of thickness $d$, at $T=0^{\circ}K$ we know that, in the limit $\Omega_{p}\rightarrow 0$ the force on the
film (the electromagnetic pressure on the film boundaries or the force between them) vanishes and the
same happens in the limit of infinite plasma frequency (perfect metal case). 
The vanishing of the force is due to the peculiar values of film reflectivity in the two limits,
it cannot stay identically zero for physical values of the reflectivity.\footnote{This can be easily understood 
by noticing that in the non-retarded regime the force on the free standing metal film is the same as the force 
between two semi-infinite plates of the same metal separated by a distance equal to the film thickness 
(see equations (\ref{smalld}) and (\ref{nomax}) in the text), which is obviously attractive and different from zero.} 
The problem of what sort of
behaviour has the force between these two limits has not been investigated: clearly, for a given film
thickness, it must reach at least one maximum of intensity. 
The questions to be answered are: (i) what is the
behaviour of the force as a function of $\Omega_{p}$, in particular at which plasma frequencies are force
maxima obtained, (ii) how do such maxima depend upon the film thickness, (iii) how is the behaviour of the
force modified when the film is deposited onto a substrate, (iv) how does the electromagnetic force on the free 
standing film compare with the film-substrate interaction, that may be responsible of adhesion and stiction
phenomena.\\
To provide answers to some of these questions, we have used the
Lifshitz approach to the electromagnetic fluctuation induced
forces to study the force on metallic films. Here we report on
some general results that can be obtained using the plasma model
model dielectric functions. Our purpose is not to achieve a
precise description of the force for specific systems, since an
accurate evaluation with an intrinsic force uncertainty of few per
cent requires a precise determination of the Drude parameters,
which are very sensitive to the sample condition
\cite{lambrecht,pirozhenko}. Rather we want to illustrate some
general trends that can be understood using a model description of
the dielectric properties.

\section{Force on isolated metallic films}
\label{ontometal} We consider metallic films of thickness $d$
ranging from $10 nm$ to a few hundred $nm$. For typical metallic
densities the electronic distribution deviates significantly from
the bulk behaviour when the size of the film is less than about
ten times the Fermi wavelength $\lambda_{F}$. Taking
$\lambda_{F}\simeq5$ \AA, the bulk description is expected to
become inaccurate when $d$ is of the order of $50$ \AA. For such
ultrathin film
quantum size effects are known to be important\cite{rogers1,rogers2,loly,lindgren1,lindgren2}.\\
\begin{figure}
\centering
\includegraphics[width=3cm,angle=0]{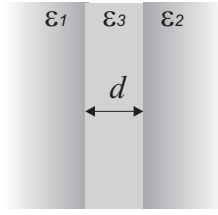}
\caption{\label{refframe3} Notation for three layers system}
\end{figure}
We adopt a local description of the electromagnetic properties of the metal based 
on a dielectric function $\epsilon(\omega)$ i.e. neglecting the wavenumber dependence of the 
dielectric response. This
local theory is expected to be less accurate in thin films than
for half space or bulk systems. Recent calculations have shown
that non local corrections to electromagnetic induced forces for
typical metallic densities are of the order of a few tenth of a
per cent, suggesting that the local theory can be appropriate in
the interpretation of the experimental data\cite{esquivel}. The
expression of the force per unit area $F$ at absolute temperature
$T$ in a configuration with two semi-infinite planar media of
dielectric functions $\epsilon_{1}(\omega)$ and
$\epsilon_{2}(\omega)$ respectively separated by a film of
thickness $d$ and dielectric function $\epsilon_{3}(\omega)$
(figure \ref{refframe3}) is given by
\begin{equation}
F=-\frac{1}{\pi\beta}\int_{0}^{\infty}k dk \sum_{n}'\bigg\{
\frac{1-Q_{TM}(i\Omega_{n})}{Q_{TM} (i\Omega_{n})}+
\frac{1-Q_{TE}(i\Omega_{n})}{Q_{TE}(i\Omega_{n})}
\bigg\}\gamma_{3} \label{exact}
\end{equation}
here $k$ is the modulus of a two dimensional wave vector parallel to the plates, 
$\beta=1/k_{B}T$, $k_{B}$ is the Boltzmann constant, $\Omega_{n}=2 \pi n / \hbar \beta$ is the
Matsubara frequency corresponding to the $n$-th thermal fluctuation mode, the prime on the summation
indicates that the $n=0$ term is given half weight. $Q_{TM}$ and $Q_{TE}$ refer to transverse magnetic (TM)
and transverse electric (TE) modes respectively and are given by
\numparts
\begin{eqnarray}
Q_{TM}(i\omega)=1&-\frac{(\epsilon_{1}\gamma_{3}-\epsilon_{3}\gamma_{1})
(\epsilon_{2}\gamma_{3}-\epsilon_{3}\gamma_{2})}{(\epsilon_{1}\gamma_{3}+\epsilon_{3}\gamma_{1})
(\epsilon_{2}\gamma_{3}+\epsilon_{3}\gamma_{2})}e^{-2d\gamma_{3}}\\
Q_{TE}(i\omega)&=1-\frac{(\gamma_{3}-\gamma_{1})
(\gamma_{3}-\gamma_{2})}{(\gamma_{3}+\gamma_{1})
(\gamma_{3}+\gamma_{2})}e^{-2d\gamma_{3}}
\end{eqnarray}
\endnumparts
with
\begin{equation}
\gamma_{i}^2=k^{2}+\frac{\Omega_{n}^{2}}{c^{2}}\epsilon_{i}(i\Omega_{n})
\label{gamma}
\end{equation}
and the dielectric functions are evaluated at the frequency $i\Omega_{n}$.\\
We performed the calculation of the force per unit area on a free
standing metal film using equation (\ref{exact}) at $T=300^{\circ}K$,
taking $\epsilon_{1}=\epsilon_{2}=1$ and
$\epsilon_{3}=1-\Omega_{3}^{2}/\omega^{2}$. We neglect for
simplicity relaxation time effects. Although they are important in
determining the infrared response of metals and the calculation we
report can be extended to a complex dielectric function, we focus
our interest mainly on general trends in the behaviour of
electromagnetic fluctuation induced forces. For realistic
calculations on specific materials relaxation time effects have to
be included and in metal films they can affect the force intensity.\\
Notice that there is a basic difference between the
electromagnetic fluctuation induced forces between two
semi-infinite metals and those on the boundaries of a film. This
is clearly seen if one considers an ideal (perfectly reflecting)
metal, corresponding to an infinite plasma frequency: the
interaction between two
semi-infinite systems is expressed by the well known Casimir force $F(d)=\hbar\pi^{2}c/240 d^{4}$, 
while for a film of finite thickness, the force on the boundaries vanishes.\\
\begin{figure}
\centering
\includegraphics[width=8cm,angle=0]{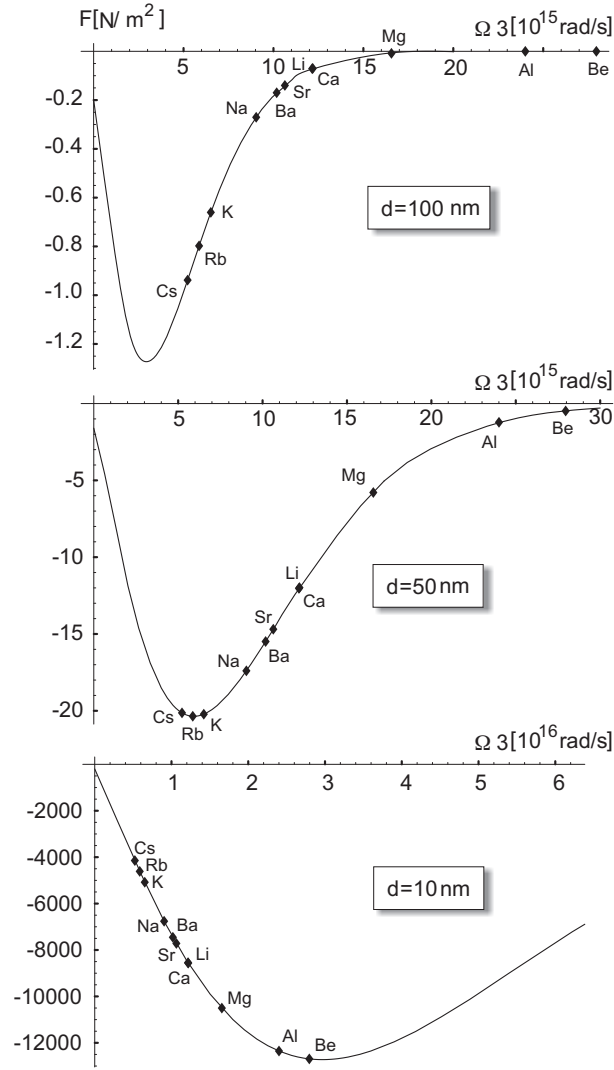}
\caption{\label{fig1} Force as a function of film plasma
frequency: the force increases on decreasing the film thickness
$d$. The calculation were performed summating the first thousand
Matsubara frequencies.}
\end{figure}
Figure \ref{fig1} displays the calculated behaviour of $F$ as a
function of the plasma frequency for films of different thickness
ranging from $10$ to $100 nm$. 
Notice that at finite temperatures the force vanishes in the large plasma frequency limit,
while for $\Omega_{p}\rightarrow 0$ there is a finite contribution from transverse magnetic modes.
This contribution is due to the $m=0$ term of the sum over Matsubara frequencies and it depends linearly upon $T$.
It is seen that the force is
attractive (it tends to contract the film) and it shows a maximum
and a tail at high plasma frequency. As expected from the general
behavior of the electromagnetic induced forces as a function of
the distance, the maximum intensity reduces as a function of $d$,
while its frequency moves to higher values. The dots in the figure
correspond to the free electron plasma frequency for sp-bonded
simple metals. This should not be seen as an accurate prediction
of the force value for metals. It indicates only that the force on
real metal films may fall on both sides of the maximum, depending
upon the film
thickness.\\
Notice that retardation effects are essential to obtain the maximum in the theoretical curve. This can be
understood by a simple calculation of the force on a free standing metal film in the van der Waals (small
$d$) regime at $T=0^{\circ}K$. In this case we have\cite{dzyaloshinskii,bergstroem}
\begin{equation}
F=-\frac{\hbar}{8\pi^{2}d^{3}}\int_{0}^{\infty}\frac{(\epsilon_{3}(i\xi)-1)^{2}}
{(\epsilon_{3}(i\xi)+1)^{2}}d\xi
\label{smalld}
\end{equation}
which leads to
\begin{equation}
F=-\frac{\hbar\Omega_{s}}{32\pi d^{3}}
\label{nomax}
\end{equation}
with $\Omega_{s}=\Omega_{3}/\sqrt{2}$ frequency of the surface plasmon. Equation (\ref{nomax}) does not show
any maximum as a function of $\Omega_{3}$. This is not surprising since the above expression is valid
under the condition that $d$ is much smaller than the plasma wavelength, therefore is appropriate in the
small plasma frequency regime only.\\
\begin{figure}
\centering
\includegraphics[width=8cm,angle=0]{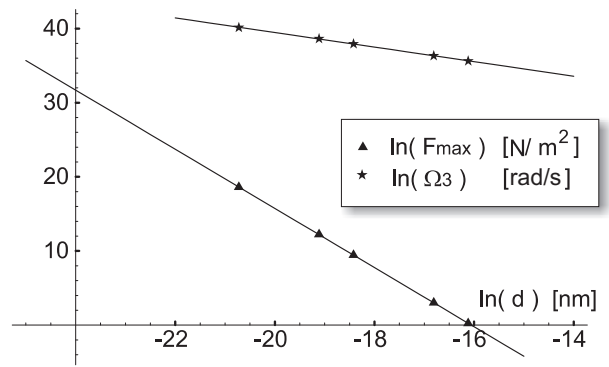}
\caption{\label{fig2} Maximum value of the force and plasma frequency at which it occurs, as a function of
distance: the fitting functions are $-64.05-3.98ln(d)$ (triangles) and $19.83-0.98ln(d)$ (stars).}
\end{figure}
The behaviour of the maximum frequency as a function of $d$ is
given in figure \ref{fig2}: it is shown that in the range of
thickness we have considered, the maximum frequency falls like
$d^{-1}$, while the intensity maximum falls as $d^{-4}$, as
expected for the interaction in the retarded regime. The behaviour
of the force maximum, that is displaced to larger values for
smaller thicknesses, can be understood by noticing that the
attraction arises from the interaction between the surface
plasmons at the two film
boundaries\cite{kampen,gerlach,intravaia}. At a given film
thickness the interaction is screened by the electron gas with
increasing efficiency as the plasma frequency increases. For large
electron density $\Omega_{3} \rightarrow\infty$, the force goes to
zero and one surface does not feel the presence of the other. The
maximum in the force results from the balance between the surface
plasmons interaction and the screening
effects. In particular for small $d$ a higher electron density is required to screen the attractive force.\\
Some interesting comments can be made on these data. The first
concerns the unsupported film stability: the force tends to shrink
the film and it has to be equilibrated by some repulsive
interaction, most likely provided by the force built up by the
valence electron rearrangement at the surfaces. Second we notice
that the force can be tuned significantly by changing the electron
density of the metal: this effect could be useful in engineering
the film properties for specific
applications.

\section{The film-ideal metal substrate interaction}
\begin{figure}
\centering
\includegraphics[width=8cm,angle=0]{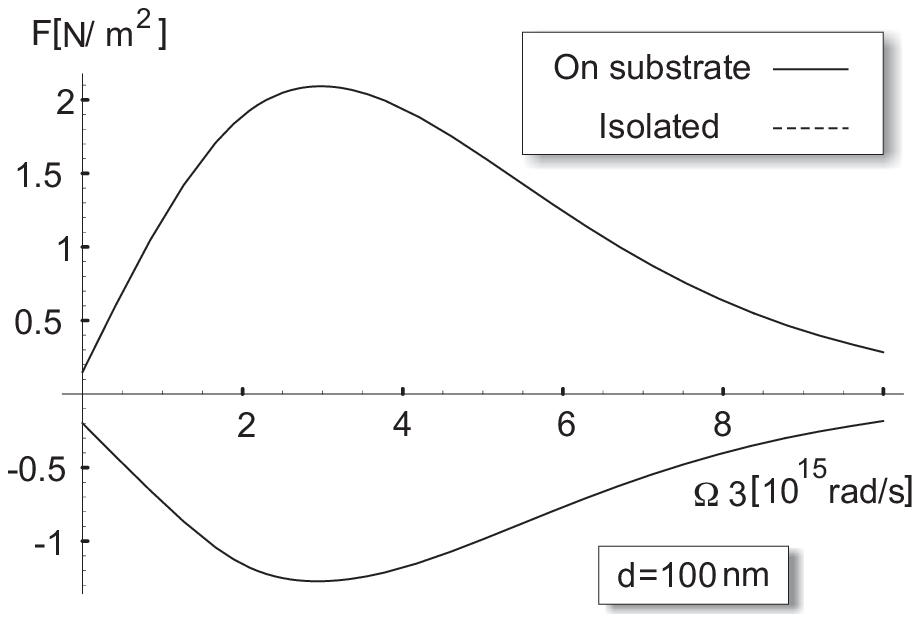}
\caption{\label{fig4} Force as a function of film plasma frequency: a change in sign occurs when the
isolated metallic film is placed on a perfectly reflecting substrate (ideal metal).}
\end{figure}

To show how these conclusions are modified when the metal film is
interacting with a substrate, we display in figure \ref{fig4} the
behaviour of the force per unit area on a film of $d=100nm$
thickness deposited onto a perfectly reflecting substrate,
(corresponding to the configuration with $\epsilon_{2}=1$ and
$\epsilon_{1}$ equal to infinity), as a function of film plasma
frequency. This is a very simplified description of a bi-metallic
interface, based on the assumption of the validity of the
continuum model, that neglects all the details of the interactions
between the atoms at the interface. It is expected to hold when
the size of the film is large compared to the interface region
(typically a few angstroms) so that the interface plays a minor
role in determining the electromagnetic force. Notice that the
force becomes repulsive and nearly double in intensity, although
it shows the same qualitative behaviour with a maximum and a long
asymmetric tail at large frequency values. It comes from the
difference between the electromagnetic force per unit area on the
substrate side and that on the vacuum side
\begin{figure}
\centering
\includegraphics[width=3cm,angle=0]{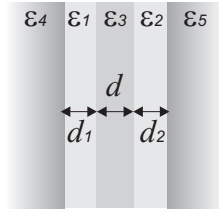}
\caption{\label{refframe5} Notation for five layers system}
\end{figure}
The behaviour of the force can be understood by noticing that at $T=0^{\circ}K$ the exact calculation in the
non-retarded limit gives the simple result:
\begin{equation}
F(d)=\frac{\hbar\Omega_p}{32\pi d^{3}}\sqrt{2}
\end{equation}
showing the change of sign and the increased force value. This result is consistent with the behaviour of 
the London dispersion forces between dissimilar materials separated by a gap, that has been reported 
since many years \cite{mahanty,israelachvili,french} . In this case the force is known to be repulsive 
when $\epsilon_{1} \gtrless \epsilon_{2} \gtrless \epsilon_{3}$ and attractive when 
$\epsilon_{1} \gtrless \epsilon_{2} \lessgtr \epsilon_{3}$ within a wide frequency range.\\
It is interesting to understand how the force between film boundaries in a 
multilayer system  is modified as a function of the film-substrate distance. For the
case of a perfectly reflecting substrate, one can determine the range of distances over which the sign of the
force changes. To this aim one has to extend equation (\ref{exact}) to a configuration with more than tree
planar media. In practice this amounts to replace the functions $Q_{TM}$ and $Q_{TE}$ by those appropriate
to a multi-layer configuration. For a five layer system the appropriate expressions were derived by Zhou and
Spruch\cite{zhou}:
\numparts
\begin{eqnarray}
Q_{TM}=Q_{TM1}Q_{TM2}\qquad Q_{TE}=Q_{TE1}Q_{TE2}\\
Q_{TM 1,TE 1}=\frac{\rho_{13}^{TM,TE}-\rho_{14}^{TM,TE}e^{-2\gamma_{1}d_{1}}}{1-\rho_{13}^{TM,TE}\rho_{14}^{TM,TE}e^{-2\gamma_{1}d_{1}}}
e^{-\gamma_{3}d}\\
Q_{TM 2,TE 2}=\frac{\rho_{23}^{TM,TE}-\rho_{25}^{TM,TE}e^{-2\gamma_{2}d_{2}}}{1-\rho_{23}^{TM,TE}\rho_{25}^{TM,TE}e^{-2\gamma_{2}d_{2}}}
e^{-\gamma_{3}d}\\
\rho_{mn}^{TE}=\frac{\gamma_{m}-\gamma_{n}}{\gamma_{m}+\gamma_{n}}\qquad
\rho_{mn}^{TM}=\frac{\gamma_{m}\epsilon_{n}-\gamma_{n}\epsilon_{m}}{\gamma_{m}\epsilon_{n}+\gamma_{n}\epsilon_{m}}
\label{5layers}
\end{eqnarray}
\endnumparts
where $\gamma_{i}$ is again given by (\ref{gamma}) and the new indexes refers to figure \ref{refframe5}.\\
For the study of substrate-metal film interaction, we take
$\epsilon_{4}$ equal to infinity,
$\epsilon_{1}=\epsilon_{2}=\epsilon_{5}=1$ while $\epsilon_{3}$ is
the metallic film dielectric function (\ref{prima}). Since the
configuration depends upon two parameters, the size $d$ of the
film and the film-substrate distance $d_{1}$, one can define the
force $F$ between the film boundaries, given by the derivative of
the free energy with respect to $d$, and the force $F'$, obtained
by deriving the free energy with respect to $d_{1}$, giving the
interaction between the film and the substrate. Figure \ref{new1}
shows the behaviour of the force $F$ on the film boundaries as a
function of the film-substrate distance for a $100 nm$ film. It is
seen that the force remains constant and attractive if the
distance $d_{1}$ is larger than the film thickness $d$; at lower
distances the force decreases until it becomes repulsive. In other
words, the film starts feeling a difference between the pressure
from the metal substrate side and the external vacuum pressure,
when the film-substrate distance is comparable with its
thickness.\\
Discussions on device stability refer usually to the interaction
between film and substrate (here we use the word substrate to
indicate a structure of much larger size than the film, it could
be a plate in a device), which gives rise to an attractive force
$F'$. To show how this interaction behaves as a function of the
ideal film-substrate distance, we have calculated $F'$ using
equations (\ref{5layers}). It turns out to be attractive for any
value of the film plasma frequency and, at distances smaller than
the film size, it is considerably more intense than the force $F$
on the film. This force is responsible of the change in sign
observed in figure \ref{fig4}: if the film is close to the
substrate, the difference between the attractive force on the film
boundaries tends to stretch the film,
causing a repulsive force between them.\\
The behaviour of the film-substrate force $F'$ in the range of
distances $d_{1}$ below the film thickness, where the substrate
effect is more significant, is illustrated by the results shown in
figure \ref{new2} for a $100nm$ film
with plasma frequency $\Omega_{3}=5\cdot 10^{15} rad/s$ and a perfectly reflecting substrate.\\
Notice that in this range of distances the force $F'$ increases
like $d_{1}^{-x}$ with $3<x<4$, (the simple $d_{1}^{-4}$ behaviour
at all distances is characteristic of the interaction between
ideal metal plates only and it is appropriate for real metals only
at large distances). At $100nm$ distance this force is
approximately $-4.8N/m^{2}$, (the Casimir force between ideal
metals at the same distance is of the order of $-10N/m^{2}$),
while the force on the film boundaries is approximately
$-1N/m^{2}$. The gray curve in the figure displays the calculated
force per unit area for a semi-infinite metal interacting with a
perfectly reflecting semi-infinite substrate. It can be seen that
it does not deviate significantly from the curve for the $100nm$
film. At higher distances the
attractive force decreases while the force on the film remains approximately constant.\\
We report in the same figure the calculated $F'$ for a $10nm$ film: in this case the force versus distance
behaviour is rather different, showing a significantly higher exponent than in the $100nm$ case ($3.52$
rather than $3.29$). Clearly this behaviour cannot be understood using arguments based on results for
semi-infinite systems: for a semi infinite metal interacting with an ideal substrate one would expect
the exponent $x$ to become closer to $3$ upon decreasing the distance. The fact that it results to be
significantly higher is a direct consequence of the finite thickness of the film. Indeed, as first pointed
out by Zhou and Sprunch\cite{zhou} higher negative exponents characterize the interaction in the presence of
film of very small thickness. An important consequence of this behaviour is that the calculated $F'$ at $10nm$ distance
(approximately $-7511.7 N/m^{2}$) is considerably higher than the force $F$ on the film boundaries (approximately $0.001 N/m^{2}$).\\
We can conclude that the interaction of a metal film with a
perfectly reflecting substrate leads to an attractive
film-substrate force and, at short distances, to a repulsive force
on the film boundaries. For $50-100nm$ thick films these forces are approximately of the same
order when the film-substrate distance is comparable with the film
size. In the low distance range ($1-10nm$) the force on the film can be
neglected and the attractive film-substrate interaction prevails
in intensity. These considerations are expected to be important
for systems, like microswitches, that consist of two conducting
electrodes, where one is fixed and the other one is able to move,
being suspended by a mechanical spring. The stability of the
system may depend upon the electromagnetic induced force acting on
the mobile film \cite{Palasantzas3,Palasantzas}.
\begin{figure}
\centering
\includegraphics[width=8cm,angle=0]{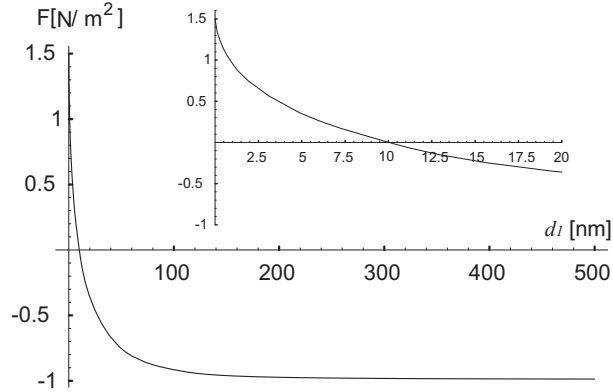}
\caption{\label{new1} Force on the film boundaries as a function of the film-substrate distance, the film plasma
frequency is $\Omega_{3}=5\cdot 10^{15}rad/s$.}
\end{figure}
\begin{figure}
\centering
\includegraphics[width=8cm,angle=0]{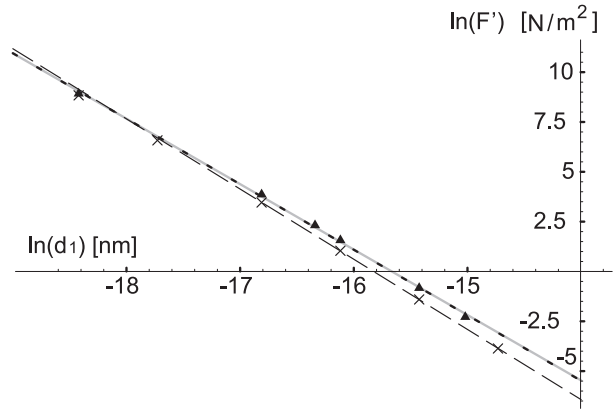}
\caption{\label{new2} Film-substrate force as a function of the
film-substrate distance between ideal metal substrate and real
metal film, film thickness $100nm$ (triangles) and $10nm$
(crosses). The fitting functions are $-51.5-3.29ln(d_{1})$ (dotted
line) and $-55.7-3.52ln(d_{1})$ (dashed line). The gray curve is
the force between two semi-infinite systems, an ideal metal and a
real metal with the same plasma frequency of the film.}
\end{figure}
\begin{figure}
\centering
\includegraphics[width=8cm,angle=0]{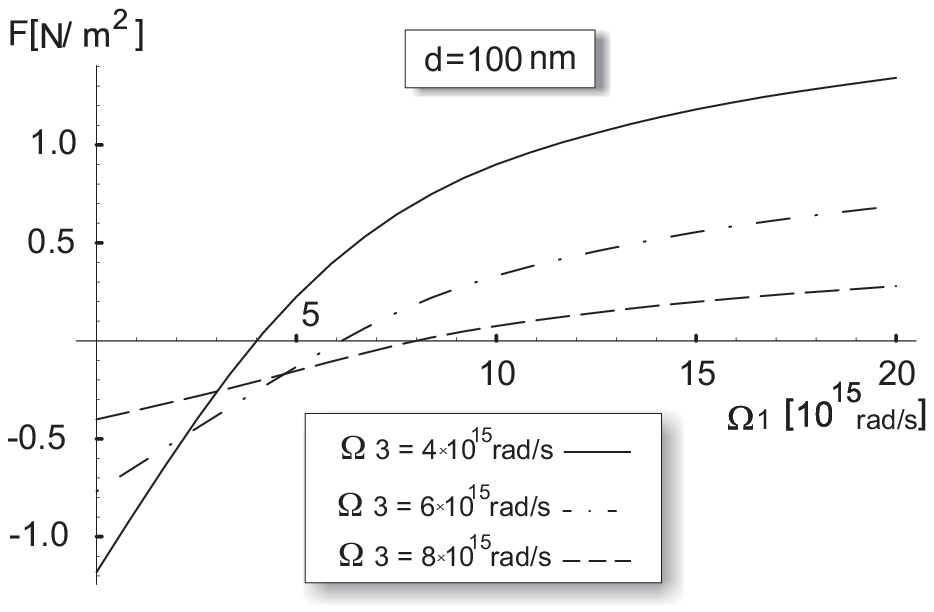}
\caption{\label{fig5a} Force on the film boundaries as a function of substrate plasma frequency calculated
for different film plasma
frequencies, in the calculation about two thousands Matsubara frequencies have been used.}
\end{figure}
\begin{figure}
\centering
\includegraphics[width=8cm,angle=0]{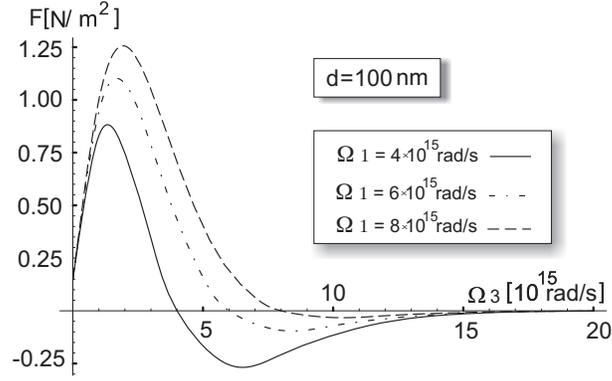}
\caption{\label{fig5b} Force on the film boundaries as a function of film plasma frequency calculated for
different substrate
frequencies, in the calculation about two thousands Matsubara frequencies have been used.}
\end{figure}

\section{The bimetallic interfaces}
\begin{figure}
\centering
\includegraphics[width=8cm,angle=0]{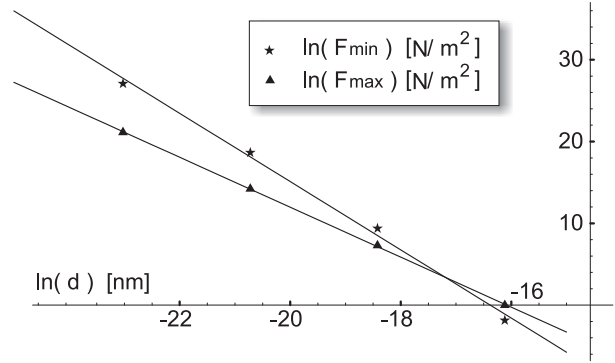}
\caption{\label{fig6} $F(\Omega_{3})$ minimum and $F(\Omega_{3})$ maximum as a function of distance $d$,
the fitting functions are respectively $-68.36-4.17ln(d)$ (stars) and $-49.10-3.05ln(d)$ (triangles), calculation were
performed at fixed $\Omega_{1}=5\times 10^{15} rad/s$}.
\end{figure}

\begin{figure}
\centering
\includegraphics[width=8cm,angle=0]{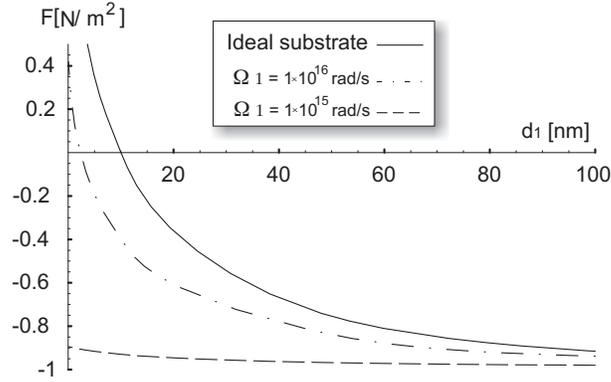}
\caption{\label{new4} Film boundaries force as a function of film-substrate distance, comparison between an
ideal substrate (continuous line), a real metal substrate with plasma frequency $10^{16}rad/s$ (dot-dash line)
and $10^{15}rad/s$ (dashed line). The film plasma frequency is $5\cdot 10^{15}rad/s$.}
\end{figure}

The situation changes if we consider a more realistic description
of the substrate. Referring to figure \ref{refframe3} this
corresponds to take $\epsilon_{1}=1-\Omega_{1}^{2}/\omega^{2}$.
Figure \ref{fig5a} shows the behaviour of the force per unit area
on a $100 nm$ metal film deposited onto various metal substrates
as a function of the substrate plasma frequency. Notice that the
force is attractive when $\Omega_{1}<\Omega_{3}$ and is repulsive
in the opposite case. For $\Omega_{1}\gg\Omega_{3}$ we get the
repulsive force corresponding to a perfectly reflecting substrate.
The change in the sign it can be easily understood by considering
the force in the small $d$ limit, i.e in the non retarded regime.
At $T=0^{\circ}K$ the force calculated from equation
(\ref{smalld}) is simply given by
\begin{equation}
F=\frac{\hbar}{32 \pi d^{3}}\frac{\Omega_{s}(\Omega_{1}^{2}-\Omega_{3}^{2})}
{\bar{\Omega}(\bar{\Omega}+\Omega_{s})}
\label{maxexist}
\end{equation}
where
\begin{equation}
\bar{\Omega}=\sqrt{(\Omega_{1}^{2}+\Omega_{3}^{2})/2}
\end{equation}
is the interface plasmon frequency obtained from the relation $\epsilon_{1}(\omega)=-\epsilon_{3}(\omega)$.
Note that $F$ shows the expected change from the repulsive to the attractive behaviour.\\
Figure \ref{fig5b} shows curves of the force on films deposited
onto different substrates as a function of the film plasma
frequencies. The curves show two extrema: on the repulsive side a
maximum, that increases in intensity and moves to higher frequency upon
increasing the substrate plasma frequency; on the attractive side
a minimum which decreases upon increasing
$\Omega_{1}$ and shifts to higher frequency values. This behaviour
is consistent with the previous conclusions concerning the ideal
substrate: as the plasma
frequency $\Omega_{1}$ increases the repulsive force on the film becomes dominant.\\
It is interesting to see how the extrema behave upon varying the
film thickness. As shown in figure \ref{fig6}, the intensity of
the repulsive maximum falls like $d^{-3}$, in the range of
distances we are considering, while for the attractive minimum it
falls approximatively as $d^{-4}$. Indeed the occurrence of the
maximum can be understood on the basis of the short distance
formula (\ref{maxexist}), which gives a $d^{-3}$ dependence of the
force, while the behaviour of the attractive part is mainly due to
retarded interactions.
These results lead to the conclusion that the electromagnetic fluctuation induced forces can give contribute
of opposite sign, and with different dependence upon the film size, to the deposited film stability.\\
As in the case of the ideal substrate, we can study the electromagnetic fluctuation induced force $F_{1}$
between the
film and the substrate as a function of the film-substrate distance. Based on the previous analysis we expect
the film-substrate force to be attractive and to lead to a repulsive or
attractive force between the film boundaries depending upon the difference between the plasma frequencies:
if $\Omega_{1}\gg\Omega_{3}$ the situation is similar to the ideal substrate case, while for
$\Omega_{1}\ll\Omega_{3}$ the force on the film is only weakly modified by the interaction. The various case
are illustrated in figure \ref{new4}.\\
\begin{figure}
\centering
\includegraphics[width=8cm,angle=0]{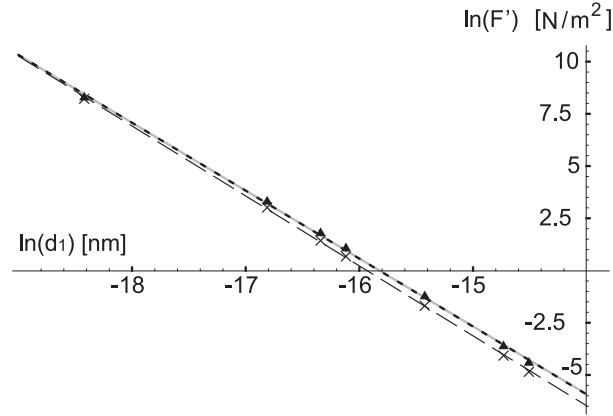}
\caption{\label{new5} Film-substrate force as a function of the film-substrate distance for real metals with
$5\cdot 10^{15}rad/s$, film thickness $100nm$ (triangles)
and $10nm$ (crosses). The fitting functions are $-51.4-3.25ln(d_{1})$ (dotted line) and
$-53.3-3.35ln(d_{1})$ (dashed line). The gray curve is the force between two semi-infinite bulks,
two real metals with the same plasma frequency.}
\end{figure}
To see how our results depend upon the film thickness we have
compared the calculated curves for film-substrate force as a
function of the distance $d_{1}$ with the electromagnetic induced
bulk-bulk interaction. We have compared the bulk-bulk interaction
with film-substrate interaction for typical values of the plasma
frequency. As shown in figure \ref{new5} the results seem not to
depend significantly upon the film thickness for $100nm$
films, while size effects become important for $d$ of the order of $10nm$.\\
It is clear, from these calculations that, in the nanometric distances range, the adoption of the simple
force expression appropriate to ideal plates is not correct. Both the sign and the intensity of the force may
result wrong, if material properties and thickness effects are not properly accounted in the theory.

\section{Discussion and Conclusions}
\label{conclusions}
We have presented a rather complete set of results based on a continuum dielectric model to illustrate trends
in the behaviour of the electromagnetic fluctuation induced forces on free-standing and supported metal
films, which allow to identify the conditions under which the force is attractive or repulsive and how it depends
upon the film thickness and the interacting substrate (plate) properties.\\
We have shown that both the sign and the intensity of the force between a film and a plate depend
upon the difference in the plasma frequencies and can be modified upon changing the carrier density. 
This is in line with the recent proposal of modulating the Casimir force between a metal and a semiconductor 
plate by illuminating the semiconducting material, i.e. by enhancing the electron plasma and creating a hole 
plasma in the semiconductor plate\cite{klimchitskaya2}. We expect that any experimental system that allows to change the 
difference in plasma frequencies can be used to modulate
the electromagnetic force.\\
An adequate description of the electronic properties in thin film
does, in general, require consideration of the changes in the
electron energy levels resulting from the confinement of the
electrons. Since these effects are observed for film thicknesses
of several nanometers, one can argue that the results of the
present paper may be significantly modified if quantum size
effects are taken into account. To clarify this point we have
calculated the dielectric permittivity of metallic films adopting
the particle in a box model\cite{wood}, in which independent
electrons are confined by a surface potential of a given length
scale $d$ along the $z$ direction, with the eigenvalue spectrum:
\begin{equation}
E_{\bf{k},n}=\frac{\hbar^{2}\bf{k}^{2}}{2 m}+E_{0}n^{2}\qquad n=1,2,3...
\end{equation}
here ${\bf k}$ is a two dimensional wavevector and $E_{0}=\hbar^{2}\pi^{2}/2 m d^{2}$. The electron
confinement leads to the quantization of the transverse component of the momentum and formation of lateral
sub-bands. The surface effects is built in the eigenstates:
\begin{equation}
\psi_{{\bf k},n}({\bf r})=\sqrt{\frac{2}{A\times d}}sin(\frac{n
\pi}{d}z)e^{i {\bf k}\cdot\bf{\rho}}
\end{equation}
where $A\times d$ is the film volume and $\bf{\rho}$ is the
positive vector in the $xy$ plane. Under such conditions the film
dielectric tensor is given by:
\begin{eqnarray}
\epsilon_{\alpha,\alpha}(\omega)=1-\frac{\Omega_{p}^{2}}{\omega^{2}}-\frac{8\pi e^{2}}{A\times d\quad m^{2}
\omega^{2}}\sum_{{\bf k},n}\sum_{{\bf k}',n'}f_{0}\times \\
\nonumber
\times (\epsilon_{\bf{k},n})\frac{
(\epsilon_{\bf{k},n}-\epsilon_{\bf{k}',n'})\vert\langle\psi_{{\bf k},n}\vert\hat{p_{\alpha}}\vert\psi_{{\bf k}',n'}
\rangle\vert^{2}}
{(\epsilon_{\bf{k},n}-\epsilon_{\bf{k}',n'})^{2}-(\hbar\omega)^{2}}
\end{eqnarray}
This expression differs from the model dielectric function in several respects: (i) it has a tensor character
with $\epsilon_{x,x}=\epsilon_{y,y}\neq\epsilon_{z,z}$, (ii) the plasma frequency $\Omega_{p}$ depends upon
the film density, which, at a fixed chemical potential at $T=0^{\circ}K$, changes as a function of the film
thickness, (iii) it accounts for transitions between lateral sub-bands (the Fermi momentum being constant the number of occupied sub-band increases upon increasing the film size). It can be easily shown that these
transitions do not affect the lateral components of the dielectric tensor. They modify the low frequency
behaviour of $\epsilon_{z,z}$. The model has been used to interpret optical and transport properties of thin
films\cite{jalochowski1,jalochowski2,jalochowski3,rogacheva}.\\
\begin{figure}
\centering
\includegraphics[width=8cm,angle=0]{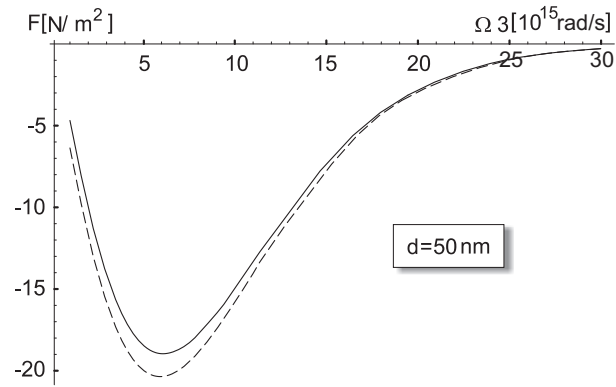}
\caption{\label{fig13} Force as a function of film plasma frequency: comparison between Drude model
(dashed line) and particle in a box model (continuous line).}
\end{figure}
We have calculated the electromagnetic induced forces on
free-standing metals films of different Fermi energy. Figure
\ref{fig13} shows the results of a calculation for a $50 nm$ film.
One can compare them with those plotted in figure \ref{fig2}. It
can be noticed that, although the value of the force is modified
by the inclusion of size effects, the behaviour as a function of
the free electron plasma frequency remains the same. Similar
results have have been obtained in other cases and will be
reported elsewhere, in a more detailed
study of quantum size effects on electromagnetic induced forces.\\
We conclude that the main trend of the results given in the present paper is not modified by quantum size
effects for thickness above $10nm$.\\
The present theory can be improved along two main lines. Inclusion
of bulk relaxation effects, both in the continuum dielectric
theory and in the particle in a box model, is expected to modify
the calculated value of the force. A more accurate description of
surface electromagnetic field, that treats the modifications to
the Fresnel optics caused by the surface, may also lead to
appreciable changes specially for film size of the order of few
nanometers.

\ack AB thanks \emph{CINECA Consorzio Interuniversitario} ({\tt
www.cineca.it}) for funding his Ph.D. fellowship.

\section*{Reference}
\bibliographystyle{unsrt}

\end{document}